\documentclass[runningheads]{llncs}
\usepackage[T1]{fontenc}

\usepackage{booktabs} 
\usepackage{color}
\usepackage{url}

\usepackage{tabularx}
\usepackage{multirow}
\usepackage[square,comma,numbers,sort&compress,sectionbib]{natbib}
\usepackage{enumerate}
\usepackage{graphicx}
\usepackage{amsmath}
\usepackage{makecell}
\usepackage{subcaption}

\newcommand{\todo}[1]{\textcolor{red}{#1}}
\newcommand{\tocheck}[1]{\textcolor{blue}{#1}}

\usepackage{fancyhdr}

\fancypagestyle{specialfooter}{%
  \fancyhf{}
  
  \fancyfoot[C]{\vspace{1cm} \hspace*{-.2\textwidth}\parbox{1.4\textwidth}{\small This is the author's version of the work. It is posted here for your personal use.}
}}

\begin{document}
\mainmatter

\title{UserSimCRS v2: Simulation-Based Evaluation for Conversational Recommender Systems}

\titlerunning{UserSimCRS v2}

\author{Nolwenn Bernard\inst{1}\orcidID{0009-0007-0565-3210} 
\and Krisztian Balog\inst{2}\orcidID{0000-0003-2762-721X}
}
\authorrunning{Bernard and Balog}

\institute{TH Köln, Köln, Germany\\ \email{nolwenn.bernard@th-koeln.de} \and
University of Stavanger, Stavanger, Norway\\ \email{krisztian.balog@uis.no}}

\maketitle
\thispagestyle{specialfooter}

\begin{abstract}
Resources for simulation-based evaluation of conversational recommender systems (CRSs) are scarce. The UserSimCRS toolkit was introduced to address this gap. In this work, we present UserSimCRS v2, a significant upgrade aligning the toolkit with state-of-the-art research. Key extensions include an enhanced agenda-based user simulator, introduction of large language model-based simulators, integration for a wider range of CRSs and datasets, and new evaluation utilities. We demonstrate these extensions in a case study.


\end{abstract}

\section{Introduction}

There is a growing interest in conversational recommender systems (CRSs), which are designed to support users in finding items corresponding to their needs and preferences through a multi-turn dialogue~\citep{Gao:2021:AIOpen}.
Despite the progress in the field, CRS evaluation remains a challenging task~\citep{Gao:2021:AIOpen,Jannach:2023:AIR,Balog:2024:FnTIR,Bernard:2025:SIGIR-AP}. Indeed, traditional evaluation methods using offline test collections do not capture the interactive nature of the systems. To address this issue, evaluation methods relying on humans are often employed, yet they suffer from significant limitations, such as their reproducibility, scalability, and cost~\citep{Kelly:2009:FnTIR}. Therefore, user simulation has recently been gaining interest to mitigate these limitations~\citep{Zhang:2020:KDD,Balog:2024:FnTIR,Breuer:2025:SIGIRForum,Schaer:2025:SIGIR,Wang:2023:EMNLP,Huang:2024:arXiv,Chen:2025:WWW,Zhu:2025:WWW,Vlachou:2025:RecSys,Zhao:2025:arXiv,Bernard:2025:SIGIR-AP}. 
While user simulation can be imperfect, it serves as a valuable preliminary step before human evaluation to refine and make a pre-selection of the best-performing systems, thereby optimizing the use of human resources~\citep{Afzali:2023:WSDM}.

Despite this need, there are few open-source resources specifically designed to perform simulation-based evaluation of CRSs. While comprehensive toolkits exist for building CRSs, such as CRSLab~\citep{Zhou:2021:ACL} and RecWizard~\citep{Zhang:2024:AAAI}, they generally lack robust and reusable resources for simulation. This creates a critical gap for researchers, who are often forced to build bespoke evaluation setups from scratch. We also observe that toolkits designed to study task-oriented dialogue systems---a generalization of CRSs to support other tasks than recommendation---include user simulators to perform training and evaluation~\citep{Lee:2019:ACL,Zhu:2020:ACL,Zhu:2023:EMNLP,Ultes:2017:ACL}. However, these are ill-suited for the recommendation domain, as they are not designed to model crucial recommendation-specific constructs, such as the assessment of recommended items based on preferences from historical interactions. Furthermore, they mostly lack support for standard recommendation benchmark datasets like ReDial~\citep{Li:2018:NIPS} and INSPIRED~\citep{Hayati:2020:EMNLP}. 

Recently, \citet{Afzali:2023:WSDM} presented UserSimCRS, a toolkit specifically tailored for simulation-based evaluation of CRSs. This original version (v1) provides the implementation of an agenda-based user simulator and basic evaluation metrics. While agenda-based simulators are foundational to the field, the research focus has decisively shifted towards user simulators based on large language models (LLMs), which represent the current state-of-the-art~\citep{Wang:2023:EMNLP,Yoon:2024:NAACL,Kim:2024:arXiv,Zhang:2025:arXiv}.
This trend is driven by their advanced natural language understanding capabilities and ability to generate more fluent, diverse, and human-like interactions, moving beyond the often rigid and predictable nature of traditional rule-based approaches, thereby promising substantially greater fidelity in modeling complex user behaviors and preferences.

However, even with a capable simulator, a significant practical challenge remains: interoperability. 
A simulation-based evaluation setup requires the simulator to communicate with the CRS.
This integration is a major barrier to adoption, as CRSs are often implemented in different frameworks and trained on disparate datasets.
Facilitating the communication between the simulator and the CRS is essential for enabling reproducible, large-scale comparative studies.

Therefore, the objective of this work is to extend UserSimCRS to provide a more holistic and modernized simulation-based CRS evaluation framework that aligns with the current state-of-the-art in research. 
These extensions include:
\begin{itemize}
    \item \emph{Enhanced agenda-based simulator}: We upgrade the classical agenda-based simulator with LLM-based components for dialogue act extraction and natural language generation, and revise the dialogue policy.
    \item \emph{LLM-based simulators}: We introduce two end-to-end LLM-based user simulators (single-prompt and dual-prompt). % using different utterance generation strategies.
    \item \emph{Integration with existing CRSs}: We introduce a new communication interface to integrate commonly used CRS models available in CRS Arena~\citep{Bernard:2025:WSDM}.
    \item \emph{Unified data framework}: We provide a unified data format with conversion and LLM-powered augmentation tools for widely-used benchmarks (ReDial~\citep{Li:2018:NIPS}, INSPIRED~\citep{Hayati:2020:EMNLP}, and IARD~\citep{Cai:2020:UMAP}).
    \item \emph{Advanced conversational quality evaluation}: We expand the toolkit's limited evaluation metrics by adding a new ``LLM-as-a-Judge'' utility along with the implementation of recently proposed user-centric utility metrics~\citep{Bernard:2025:SIGIR-AP}.
\end{itemize}
We showcase these extensions through a case study in the movie recommendation domain. Specifically, we perform simulation-based evaluation of a selection of CRSs using the different types of user simulators and datasets now supported in UserSimCRS v2. The most recent version of UserSimCRS is available at \url{https://github.com/iai-group/UserSimCRS}.

\section{Related Work}
\label{sec:related}

User simulation for the evaluation of interactive information access systems has seen a renewed interest in recent years~\citep{Balog:2024:FnTIR,Breuer:2025:SIGIRForum,Schaer:2025:SIGIR}.
Our work is situated within this trend, focusing specifically on conversational recommender systems.

Developing a CRS is a technically challenging task. Therefore, various toolkits have been proposed to lower the technical barrier. % to implement new and reusable CRSs. 
For example, CRSLab~\citep{Zhou:2021:ACL} toolkit supports development with deep neural networks, while FORCE~\citep{Quan:2022:AAAI} and RecWizard~\citep{Zhang:2024:AAAI} support rule-based and LLM-based CRSs, respectively. These toolkits come with a user interface allowing direct interaction with the CRS and provide limited resources for evaluation. Moreover, offline evaluation commonly focuses on specific components of the CRS, e.g., evaluation of the recommender with mean reciprocal rank or natural language generator with BLEU~\citep{Zhou:2021:ACL}.
This contrasts with platforms dedicated to large-scale \emph{human} evaluation, such as CRS Arena~\citep{Bernard:2025:WSDM}, which benchmarks CRSs by having users judge pairwise battles between anonymous systems.

\begin{figure}[t]
    \centering
    \begin{subfigure}{\textwidth}
        \centering
        \includegraphics[width=\textwidth]{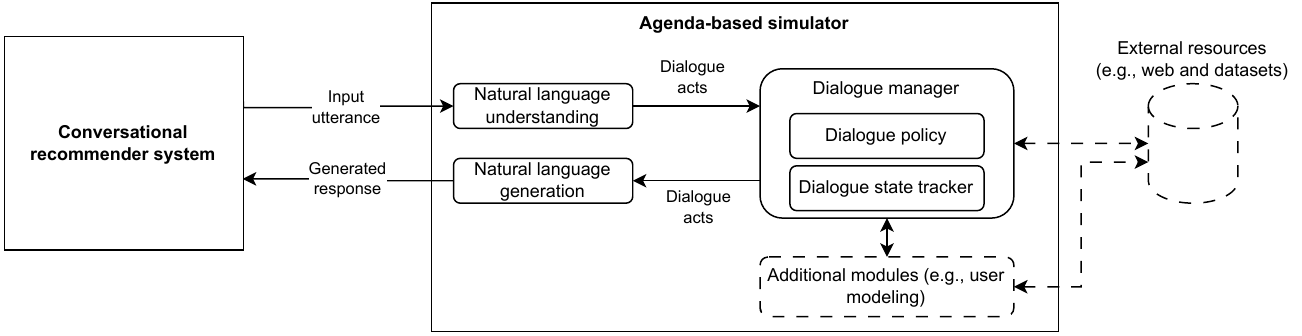}
        \caption{Agenda-based architecture.}
        \label{fig:architectures:abus}
    \end{subfigure}
    \begin{subfigure}{\textwidth}
        \centering
        \includegraphics[scale=0.6]{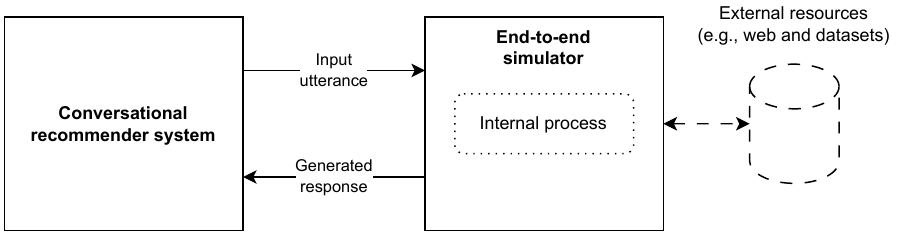}
        \caption{End-to-end architecture.}
        \label{fig:architectures:ete}
    \end{subfigure}
    \caption{Agenda-based (Top) and end-to-end (Bottom) architectures for user simulators. The dashed lines represent optional components and data flows.}
    \label{fig:architectures}
\end{figure}

User simulators are commonly built using either a modular or end-to-end architecture. In a modular architecture, different components with specific roles interact with each other to process the incoming utterance and generate a response. For example, the established agenda-based simulator~\citep{Schatzmann:2007:NAACL} typically comprises three components, as illustrated in Fig.~\ref{fig:architectures:abus}, to transform the incoming utterance to a structured representation (i.e., dialogue acts), decide on the next action to take, and then transform it to a natural response.
While in an end-to-end architecture, the user simulator processes the incoming utterance and directly generates the response in natural text (Fig.~\ref{fig:architectures:ete}), e.g.,~\citep{Wang:2023:EMNLP,Kim:2024:ACL}. 
This architecture is commonly used for user simulators using a (deep) neural network or a large language model.

Simulation-based evaluation of CRSs is still in its early stages, in part due to the complexity of the task and the limited availability of resources~\citep{Balog:2024:FnTIR}. However, some initiatives have recently been proposed to reduce this gap and make simulation-based evaluation more accessible. These include the evaluation methodologies iEvaLM~\citep{Wang:2023:EMNLP} and CONCEPT~\citep{Huang:2024:arXiv} (built on top of iEvaLM), which share similar objectives to UserSimCRS. They facilitate simulation-based evaluation of CRSs using user simulators based on LLMs. iEvaLM considers objective (e.g., recall) and subjective (e.g., persuasiveness) metrics, while CONCEPT uses system- and user-centric factors, such as reliability, cooperation, and social awareness, to assess a CRS's performance. However, we note that their implementation exclusively supports CRSs implemented in CRSLab~\citep{Zhou:2021:ACL} and trained on either ReDial~\citep{Li:2018:NIPS} or OpenDialKG~\citep{Moon:2019:ACL}. 

The use of LLMs as surrogates for humans in evaluation, often termed ``LLM-as-a-Judge''~\citep{Zheng:2023:NeurIPS}, has become a prevalent trend. This approach is now widely applied across diverse tasks, including relevance assessment in information retrieval~\citep{Faggioli:2023:ICTIR,Thomas:2024:SIGIR,Dietz:2025:ICTIR} and multi-faceted evaluation of CRSs based on detailed scoring rubrics~\citep{Huang:2024:arXiv,Wang:2023:EMNLP}. Despite its widespread adoption, significant open questions remain, including susceptibility to biases and manipulation~\citep{Balog:2025:SIGIR,Wang:2024:ACL,Alaofi:2024:SIGIRAP}.

\section{UserSimCRS v1}
\label{sec:usersimcrs}

UserSimCRS~\citep{Afzali:2023:WSDM} is a toolkit facilitating simulation-based evaluation of conversational recommender systems. It is built on top of DialogueKit,\footnote{\url{https://github.com/iai-group/DialogueKit}} a library providing dialogue management and evaluation functionalities, and comprises three main modules: 
\begin{itemize}
    \item \emph{Agenda-based user simulator}: The implementation of a modular agenda-based user simulator, which contains the logic to generate user responses based on the dialogue history and the user's model.
    \item \emph{User modeling}: Allows the definition of user models covering their preferences, the context, and their persona.
    \item \emph{Item collection}: Provides a unified representation of the possible items to recommend.
\end{itemize}
The evaluation process consists of four main steps. First, the experimental setup is defined, including the interaction model, the user population(s), and the evaluation metrics. In the subsequent steps, the user simulator is trained and a set of conversations is generated. Finally, the evaluation metrics (e.g., success rate and average number of turns) are computed. This method permits the relative comparison of different CRSs under the same conditions.

While the toolkit is a valuable asset for the community, we believe that the following limitations hinder its widespread adoption.  
\begin{itemize}
    \item \emph{Limited selection of user simulators}: The toolkit exclusively provides an agenda-based simulator. While foundational, this approach is becoming outdated and lacks the generative flexibility of LLM-based simulators, which are now prevalent and represent the current state-of-the-art research direction.
    \item \emph{Limited components for agenda-based simulator}: The original implementation of the agenda-based user simulator is modular; however, it relies on a supervised model for natural language understanding and a template-based model for response generation. This limits the diversity of the responses and requires training data to be effective.
    \item \emph{Lack of explicitly defined information need}: The preferences and values associated with the elicited slots are randomly sampled, which can lead to incoherent conversations. In theory, the user simulator should be initialized with an information need that guides the conversation and the user responses~\citep{Schatzmann:2007:NAACL}.
    \item \emph{Lack of support for benchmark datasets}: The toolkit provides a dataset based on the IAI MovieBot agent~\citep{Habib:2020:CIKM} to create a user simulator, but it is not widely used in the field. Hence, the preparation of popular benchmark datasets is left to the toolkit users which represents a significant burden. 
    \item \emph{Limited selection of evaluation measures}: Evaluation functionalities in DialogueKit are limited to the computation of user satisfaction and average number of turns. The computation of other metrics, especially those related to recommendation and conversational quality, is the responsibility of the toolkit user (i.e., the CRS developer).
\end{itemize}
The objective of this work is to extend UserSimCRS to address these limitations.
\section{UserSimCRS v2 Extensions}
\label{sec:v2}

\begin{figure}[t]
    \centering
    \includegraphics[keepaspectratio, scale=0.75]{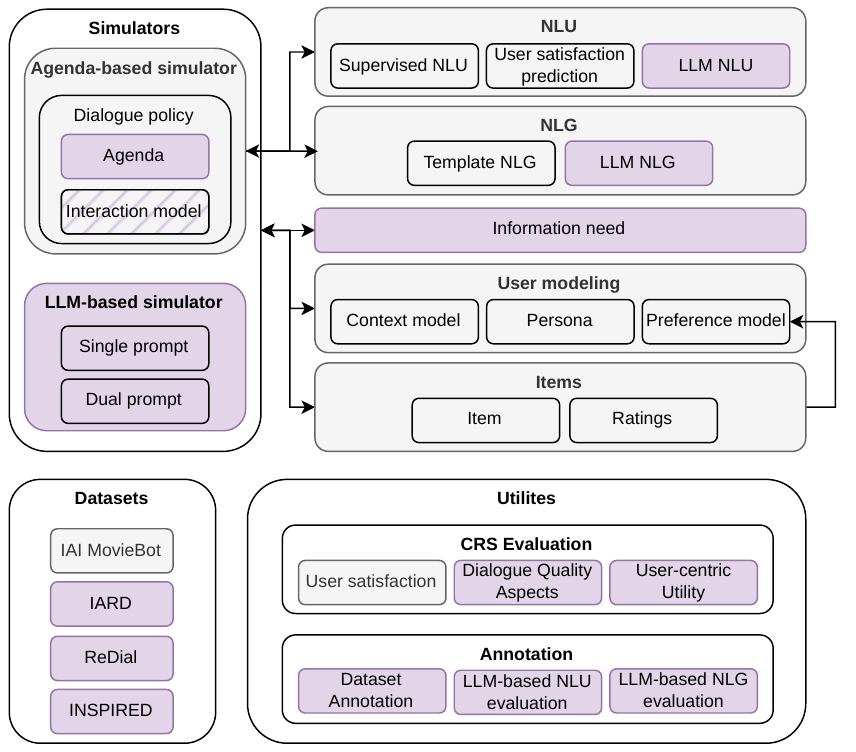}
    \caption{Overview of UserSimCRS v2 architecture. Grey components are inherited from \citep{Afzali:2023:WSDM}, while hashed and purple components correspond to updated or added components, respectively.}
    \label{fig:architecture}
\end{figure}

The underlying objective of UserSimCRS v2 remains the same as its predecessor: to facilitate the comparison of different conversational recommender systems via simulation-based evaluation. Additionally, it can be used to compare and investigate different user simulators. 
In this section, we present the extensions introduced in UserSimCRS v2, which are illustrated in Fig.~\ref{fig:architecture}. Specifically, we describe the integration of existing benchmark datasets (Section~\ref{sec:v2:datasets}) the improvements to the agenda-based user simulator (Section~\ref{sec:v2:agenda}), the introduction of LLM-based user simulators (Section~\ref{sec:v2:llm}), the integration of existing CRSs (Section~\ref{sec:v2:crss}), and the addition of evaluation utilities and metrics to assess conversation quality and user utility (Section~\ref{sec:v2:eval}).

\subsection{Unified Data Format and Benchmark Datasets}
\label{sec:v2:datasets}

\begin{table}[t]
    \centering
    \caption{Summary of datasets added in UserSimCRS v2. $^*$ indicates augmented version of the original dataset.}
    \label{tab:datasets}
    \begin{tabular}{l|l|c|l}
        \toprule
        Dataset & Domain & \# Dialogues & Dialogue annotations \\
        \midrule
        IARD$^*$~\citep{Cai:2020:UMAP} & Movie & 77 & Dialogue acts, recommendations accepted \\
        INSPIRED~\citep{Hayati:2020:EMNLP} & Movie & 1,001 & Social strategies, entity mentioned, item feedback \\ 
        ReDial~\citep{Li:2018:NIPS} & Movie & 10,006 & Movies references, item feedback \\
        \bottomrule
    \end{tabular}
\end{table}

UserSimCRS v1 provides an initial dataset to create a user simulator, built from interactions with IAI MovieBot~\citep{Habib:2020:CIKM}; however, this dataset is not widely used in the field. In UserSimCRS v2, we aim to reduce the barrier to adoption by providing a unified data format and conversion tools for the widely-used conversational datasets INSPIRED~\citep{Hayati:2020:EMNLP} and ReDial~\citep{Li:2018:NIPS}. Additionally, we provide an augmented version of IARD~\citep{Cai:2020:UMAP}, a subset of ReDial~\citep{Li:2018:NIPS} with intent annotations; see Table~\ref{tab:datasets} for a summary.

The proposed unified data format is based on the format used in DialogueKit, where utterances are annotated with dialogue acts. A dialogue act comprises an intent and optionally a set of associated slot-value pairs. For example, the dialogue acts \texttt{Elicit(size)} and \texttt{Elicit(color)} are present in the following CRS utterance: \emph{``What size and color do you prefer for your new skates?''} This allows for a refined representation of utterances and to support more complex interactions than UserSimCRS v1, which only considers a single intent and a list of slot-value pairs per utterance.
Furthermore, we introduce the notion of an \emph{information need} that consists of a set of constraints $C$ and requests $R$ as in~\citep{Schatzmann:2007:NAACL}, in addition to target items (i.e., what the user is looking for). For example, the information need of a user looking for the song \emph{Happy} by \emph{Pharrell Williams} and who wants to know the release year can be represented as follows:
\begin{align*}
    C &= \left[ \begin{array}{lll}
        \mathrm{genre} & = & \mathrm{pop} \\
        \mathrm{artist} & = & \mathrm{Pharrell\ Williams} \\
    \end{array} \right] \\
    R &= \left[ \begin{array}{lll}
        \mathrm{release\_year} & = & \mathrm{?} \\
    \end{array} \right] \\
    \mathrm{target} & = \mathrm{Happy}
\end{align*}
This information need guides the conversation and the user responses, in addition to informing the user's preferences and persona. In our example, a response to the CRS utterance eliciting the looked-for genre could be: \emph{``I like pop music, what do you recommend?''}

We note that these benchmark datasets do not necessarily contain all the annotations required to build and train certain user simulators. For example, the utterances in ReDial are not annotated with dialogue acts which are needed to train a user simulator or some components of it (e.g., a supervised natural language understanding model). In such cases, we propose to perform automatic data augmentation.
Previous work~\citep{Su:2025:CP,Zhangwenbo:2025:ACL} argues that LLMs are an acceptable alternative to human annotators in data-scarce scenarios.
Therefore, in UserSimCRS v2, we provide tools relying on pre-trained LLMs for the tasks of dialogue act annotation and information need extraction. However, when using these tools, the biases and limitations of LLMs should be acknowledged and sanity checks are recommended.

\subsection{Enhanced Agenda-based User Simulator}
\label{sec:v2:agenda}

UserSimCRS v1 implements the agenda-based user simulator as a modular system (Fig.~\ref{fig:architectures:abus}) with components for natural language understanding (NLU), dialogue policy,\footnote{Referred to as response generation in v1.} and natural language generation (NLG).
In UserSimCRS v2, we introduce new NLU and NLG components, based on an LLM. They offer ready-to-use alternatives to the original supervised components, which require annotated training data. 
Furthermore, we revised the dialogue policy component to be guided by the information need and to better reflect the natural flow of recommendation dialogues.

\paragraph{Natural Language Understanding.}
The objective of the NLU component is to extract dialogue acts from incoming CRS utterances. 
In UserSimCRS v2, we implement a LLM-based NLU component which is  initialized with a list of possible intents and slots for the dialogue acts and a prompt that includes a placeholder for the incoming CRS utterance. Dialogue acts extraction is performed by sending the prompt with a given utterance to an LLM and parsing the output based on the following expected format: \texttt{intent1(slot="value",slot,\ ...)|intent2()}.
We provide a default prompt, built using a few-shot approach with examples taken from the augmented IARD dataset, to extract dialogue acts in the movie domain, along with its corresponding configuration.   

\paragraph{Dialogue Policy.}
The dialogue policy determines the dialogue acts of the next user utterance based on the current agenda and dialogue state. In the agenda-based simulator, this operation is handled by the interaction model. For UserSimCRS v2, the interaction model is revised to align with the new data format that uses dialogue acts and an information need to guide the conversation.
In particular, the agenda is now initialized based on the information need, including a disclosure dialogue act for each constraint and an inquiry dialogue act for each request.
Furthermore, we modify the agenda update process to better reflect the natural flow of recommendation dialogues, drawing inspiration from \citet{Lyu:2021:WWW}. They observe that an initial elicitation stage typically precedes the first recommendation, after which the user makes inquiries and critiques on the recommended items. Accordingly, our new process explicitly handles four cases: (1) elicitation, (2) recommendation, (3) inquiry, and (4) other. 
A different type of dialogue act is pushed to the agenda for the first three cases; for example, a disclosure dialogue act is added to the agenda when the CRS elicits information. In the last case, the next dialogue act is either taken from the agenda (if coherent with the current dialogue state) or sampled based on historical data. Note that all the slot-value pairs in the dialogue acts are derived from the information need or from the user's preferences if not present in the information need. 

\paragraph{Natural Language Generation.}
The NLG component is responsible for generating human-like textual utterance based on the dialogue acts provided by the dialogue policy component.
In UserSimCRS v2, we introduce an LLM-based NLG component.
It aims to increase the diversity of the generated responses compared to the template-based NLG component from UserSimCRS v1.
To generate an utterance, this component uses a prompt that includes placeholders for dialogue acts and, optionally, additional annotations (e.g., emotion).  
We also provide a default prompt, following a few-shot approach with examples from the augmented IARD dataset, for the movie domain, alongside its corresponding configuration.

\subsection{Large Language Model-based User Simulator}
\label{sec:v2:llm}

The initial version of UserSimCRS only implements agenda-based user simulation. However, other simulation approaches exist, and recently, we observe a growing interest in LLM-based user simulation~\citep{Zhang:2025:arXiv}.
To align with this trend and increase the versatility of the toolkit, UserSimCRS v2 introduces two LLM-based user simulators following different utterance generation strategies. %, which we describe below.

\paragraph{Single-Prompt Simulator.}
The single-prompt user simulator produces the next user utterance in an end-to-end manner given a single prompt. The prompt, inspired by \citet{Terragni:2023:arXiv}, includes the task description, an optional persona, the information need, and the current conversation history. Unlike \citet{Terragni:2023:arXiv}, our implementation follows a zero-shot approach, i.e., it does not include any examples of conversations.
Nevertheless, one could include examples in the task description to follow a few-shot approach. 

\paragraph{Dual-Prompt Simulator.}
The dual-prompt user simulator extends the single-prompt simulator by first using a separate prompt to decide whether to continue the conversation  (the ``stopping prompt''). If the decision is to continue, it generates the next utterance using the main generation prompt (identical to the one used by the single-prompt simulator); otherwise, it sends a default utterance to stop the conversation. 
The stopping prompt is designed similarly to the main generation prompt, including the task description (re-framed to focus on taking a binary decision), an optional persona, and the current conversation history.

\subsection{Integration with Large Language Models}
\label{sec:v2:llms}

UserSimCRS v2 introduces several configurable components and user simulators relying on an LLM, all of which assumes the LLM employed is hosted on a server. To manage interactions with the server, all LLM-based modules use a common interface. UserSimCRS v2 provides two initial implementations of this interface: one for OpenAI\footnote{\url{https://openai.com/api/}} and one for an Ollama\footnote{\url{https://ollama.com/}} server. This design makes the toolkit highly extensible, as additional interfaces for other LLM backends can be easily added through inheritance.

\subsection{Integration with Existing CRSs}
\label{sec:v2:crss}

By design, UserSimCRS considers the CRS as a ``black box,'' i.e., it does not assume access to its source code or inner workings. The only requirement is to have an interface to interact with the CRS; in practice, this is done using the DialogueKit library. 
The original version of UserSimCRS provides an interface to communicate with IAI MovieBot~\citep{Habib:2020:CIKM}, a CRS for movie recommendation.
In UserSimCRS v2, we introduce a new interface to interact with CRSs available in CRS Arena~\citep{Bernard:2025:WSDM}, such as KBRD~\citep{Chen:2019:EMNLP}, BARCOR~\citep{Wang:2022:arXiv}, and ChatCRS~\citep{Wang:2023:EMNLP}. This integration reduces the technical burden of using the toolkit, while facilitating experimentation with more diverse and commonly used CRSs.

\subsection{Advanced Evaluation Metrics}
\label{sec:v2:eval}

Evaluation in UserSimCRS v1 is limited to user satisfaction and average number of turns. However, there are additional measures to assess the performance of CRSs, such as those related to recommendation accuracy and linguistic abilities~\citep{Jannach:2023:AIR}. In UserSimCRS v2, we provide new evaluation utilities to assess conversation quality using a LLM-based evaluator (``LLM-as-a-Judge''). We also implement novel metrics that measure utility from a user-centric perspective~\citep{Bernard:2025:SIGIR-AP}.

\paragraph{Conversation Quality Evaluation with LLM-as-a-Judge.}
Conversation quality is assessed with respect to five aspects inspired by \citet{Huang:2024:arXiv}:
(1) \emph{Recommendation relevance} measures how closely
the recommended items align with the user's preferences and needs;
(2) \emph{Communication style} corresponds to the conciseness and clarity of the responses;
(3) \emph{Fluency} is the degree of naturalness of the responses compared to human-generated responses;
(4) \emph{Conversational flow} assesses the coherence and consistency of the conversation; and
(5) \emph{Overall satisfaction} encapsulates the user's holistic experience.
Each aspect is assigned a score between 1 and 5, with each score defined in a grading rubric.\footnote{The grading rubrics are available at: \url{https://github.com/iai-group/UserSimCRS/blob/main/scripts/evaluation/rubrics/quality_rubrics.py}}
For example, for \emph{recommendation relevance} a score of 1 represents irrelevant recommendations and a score of 5 represents highly relevant recommendations.
We acknowledge that there are open questions regarding the use of LLM-based evaluators, as correlation with human judgments varies across different studies~\citep{Dietz:2025:ICTIR}.

\paragraph{User-centric Utility Metrics}.
\citet{Bernard:2025:SIGIR-AP} propose metrics to assess the utility of CRSs from a user-centric perspective. We implement Successful Recommendation Round Ratio (SRRR) and Reward-per-Dialogue-Length (RDL), in addition to success rate. SRRR measures the proportion of recommendation rounds resulting in an acceptance, while RDL divides the number of accepted recommendations by dialogue length. The computation of these metrics relies on the underlying intent of user utterances in a conversation, that are annotated either manually or automatically using an NLU model.

\section{Use Case: Movie Recommendation}
\label{sec:example}

To demonstrate the use of different user simulators and CRSs supported in UserSimCRS v2, we present a case study in the movie recommendation domain.

\paragraph{User Simulators.} We instantiate the different types of user simulators supported in UserSimCRS v2---agenda-based (ABUS), single/dual-prompt LLM-based (LLM-SP/DP)---using different datasets.
Depending on the type of the simulator and the dataset, we prepare the item collection, define an interaction model, create prompts, and annotate a sample of dialogues with dialogue acts.

\paragraph{Conversational Recommender Systems.} We consider four conversational recommender systems from the CRS Arena~\citep{Bernard:2025:WSDM}: BARCOR\_OpenDialKG, BARCOR\_ReDial, KBRD\_ReDial, and UniCRS\_OpenDialKG. Additionally, we include IAI MovieBot, as it was already supported in UserSimCRS v1.

\begin{table*}[t]
\caption{Evaluation of selected CRSs with different user simulators. Results reported correspond to average scores over 100 synthetic dialogues.}
\label{tab:results}
\scriptsize
\centering
\begin{tabular}{l|ccc|ccc|ccc}
    \toprule
    \multirow{2}{*}{\textbf{CRS}} & \multicolumn{3}{c|}{\textbf{User satisfaction}} & \multicolumn{3}{c|}{\textbf{Fluency}} & \multicolumn{3}{c}{\textbf{Rec. relevance}} \\
    & ABUS & LLM-SP & LLM-DP & ABUS & LLM-SP & LLM-DP & ABUS & LLM-SP & LLM-DP \\
    \midrule
    \multicolumn{10}{l}{\emph{Dataset: MovieBot}} \\
    \midrule
    BARCOR\_OpenDialKG & 1.9 \tiny{$\pm 0.6$} & 1.9 \tiny{$\pm 0.6$} & 1.9 \tiny{$\pm 0.6$} & 2.8 \tiny{$\pm 0.5$} & 2.6 \tiny{$\pm 0.6$} & 3.0 \tiny{$\pm 0.5$} & 1.4 \tiny{$\pm 0.6$} & 1.5 \tiny{$\pm 0.9$} & 2.0 \tiny{$\pm 0.9$} \\
    BARCOR\_ReDial & 2.0 \tiny{$\pm 0.5$} & 1.8 \tiny{$\pm 0.5$} & 2.0 \tiny{$\pm 0.6$} & 3.3 \tiny{$\pm 0.8$} & 2.6 \tiny{$\pm 0.6$} & 2.9 \tiny{$\pm 0.6$} & 2.4 \tiny{$\pm 1.6$} & 1.5 \tiny{$\pm 0.9$} & 1.9 \tiny{$\pm 0.9$} \\
    KBRD\_ReDial & 1.9 \tiny{$\pm 0.4$} & 1.9 \tiny{$\pm 0.4$} & 1.9 \tiny{$\pm 0.4$} & 2.5 \tiny{$\pm 0.6$} & 2.4 \tiny{$\pm 0.6$} & 2.6 \tiny{$\pm 0.6$} & 1.2 \tiny{$\pm 0.6$} & 1.3 \tiny{$\pm 0.7$} & 1.5 \tiny{$\pm 0.8$} \\
    UniCRS\_OpenDialKG & 2.3 \tiny{$\pm 0.8$} & 1.8 \tiny{$\pm 0.6$} & 1.9 \tiny{$\pm 0.6$} & 2.4 \tiny{$\pm 0.5$} & 2.5 \tiny{$\pm 0.6$} & 2.8 \tiny{$\pm 0.5$} & 1.0 \tiny{$\pm 0.2$} & 1.5 \tiny{$\pm 1.0$} & 1.5 \tiny{$\pm 0.7$} \\
    IAI MovieBot & 1.5 \tiny{$\pm 0.7$} & 1.7 \tiny{$\pm 0.8$} & 2.5 \tiny{$\pm 0.7$} & 3.4 \tiny{$\pm 0.6$} & 3.1 \tiny{$\pm 0.7$} & 3.3 \tiny{$\pm 1.0$} & 2.1 \tiny{$\pm 1.1$} & 2.3 \tiny{$\pm 1.4$} & 2.9 \tiny{$\pm 1.7$} \\
    \midrule
    \multicolumn{10}{l}{\emph{Dataset: INSPIRED}} \\
    \midrule
    BARCOR\_OpenDialKG & 2.2 \tiny{$\pm 0.7$} & 2.0 \tiny{$\pm 0.6$} & 2.1 \tiny{$\pm 0.6$} & 2.6 \tiny{$\pm 0.6$} & 2.7 \tiny{$\pm 0.5$} & 3.0 \tiny{$\pm 0.6$} & 1.6 \tiny{$\pm 0.6$} & 1.4 \tiny{$\pm 0.6$} & 1.9 \tiny{$\pm 1.0$} \\
    BARCOR\_ReDial & 2.1 \tiny{$\pm 0.5$} & 2.0 \tiny{$\pm 0.5$} & 2.0 \tiny{$\pm 0.5$} & 3.5 \tiny{$\pm 0.5$} & 2.6 \tiny{$\pm 0.6$} & 2.9 \tiny{$\pm 0.6$} & 2.3 \tiny{$\pm 0.8$} & 1.4 \tiny{$\pm 0.9$} & 1.6 \tiny{$\pm 0.9$} \\
    KBRD\_ReDial & 2.0 \tiny{$\pm 0.3$} & 2.0 \tiny{$\pm 0.5$} & 2.1 \tiny{$\pm 0.6$} & 3.3 \tiny{$\pm 0.6$} & 2.3 \tiny{$\pm 0.5$} & 2.6 \tiny{$\pm 0.6$} & 1.9 \tiny{$\pm 0.9$} & 1.2 \tiny{$\pm 0.6$} & 1.4 \tiny{$\pm 0.7$} \\
    UniCRS\_OpenDialKG & 1.9 \tiny{$\pm 0.8$} & 2.1 \tiny{$\pm 0.6$} & 2.1 \tiny{$\pm 0.6$} & 2.9 \tiny{$\pm 0.5$} & 2.4 \tiny{$\pm 0.5$} & 2.7 \tiny{$\pm 0.5$} & 1.3 \tiny{$\pm 0.6$} & 1.1 \tiny{$\pm 0.4$} & 1.5 \tiny{$\pm 0.7$} \\
    IAI MovieBot & 1.4 \tiny{$\pm 0.8$} & 2.2 \tiny{$\pm 0.8$} & 2.2 \tiny{$\pm 0.8$} & 3.3 \tiny{$\pm 0.5$} & 2.9 \tiny{$\pm 1.0$} & 3.3 \tiny{$\pm 1.0$} & 2.1 \tiny{$\pm 0.8$} & 2.7 \tiny{$\pm 1.9$} & 3.1 \tiny{$\pm 1.8$} \\
    \midrule
    \multicolumn{10}{l}{\emph{Dataset: ReDial}} \\
    \midrule
    BARCOR\_OpenDialKG & 2.2 \tiny{$\pm 0.8$} & 1.9 \tiny{$\pm 0.8$} & 2.1 \tiny{$\pm 0.7$} & 3.0 \tiny{$\pm 0.6$} & 2.6 \tiny{$\pm 0.5$} & 3.0 \tiny{$\pm 0.5$} & 1.9 \tiny{$\pm 0.8$} & 1.4 \tiny{$\pm 0.6$} & 1.8 \tiny{$\pm 0.9$} \\
    BARCOR\_ReDial & 2.0 \tiny{$\pm 0.1$} & 2.2 \tiny{$\pm 0.6$} & 2.3 \tiny{$\pm 0.8$} & 3.4 \tiny{$\pm 0.6$} & 2.4 \tiny{$\pm 0.5$} & 2.4 \tiny{$\pm 0.6$} & 2.3 \tiny{$\pm 0.9$} & 1.2 \tiny{$\pm 0.6$} & 1.3 \tiny{$\pm 0.6$} \\
    KBRD\_ReDial & 1.9 \tiny{$\pm 0.6$} & 2.1 \tiny{$\pm 0.5$} & 2.1 \tiny{$\pm 0.8$} & 3.0 \tiny{$\pm 0.5$} & 2.1 \tiny{$\pm 0.4$} & 2.3 \tiny{$\pm 0.5$} & 1.7 \tiny{$\pm 0.8$} & 1.0 \tiny{$\pm 0.1$} & 1.2 \tiny{$\pm 0.6$} \\
    UniCRS\_OpenDialKG & 2.0 \tiny{$\pm 0.8$} & 1.9 \tiny{$\pm 0.8$} & 1.9 \tiny{$\pm 0.8$} & 2.7 \tiny{$\pm 0.5$} & 2.3 \tiny{$\pm 0.5$} & 2.8 \tiny{$\pm 0.6$} & 1.3 \tiny{$\pm 0.5$} & 1.1 \tiny{$\pm 0.3$} & 1.5 \tiny{$\pm 0.9$} \\
    IAI MovieBot & 2.0 \tiny{$\pm 0.2$} & 2.3 \tiny{$\pm 0.9$} & 2.2 \tiny{$\pm 0.9$} & 4.0 \tiny{$\pm 0.2$} & 2.9 \tiny{$\pm 1.0$} & 3.2 \tiny{$\pm 1.0$} & 4.7 \tiny{$\pm 0.8$} & 2.4 \tiny{$\pm 1.7$} & 2.8 \tiny{$\pm 1.6$} \\
    \bottomrule
\end{tabular}
\end{table*}

\paragraph{Evaluation.} We evaluate the selected CRSs using the different user simulators. This process consists of generating 100 synthetic dialogues for each pair of user simulator and CRS, and computing evaluation metrics over these dialogues. In Table~\ref{tab:results}, we report user satisfaction, fluency, and recommendation relevance.\footnote{Configurations and synthetic dialogues are available in the repository.} 
The results confirm that, in general, CRS performance remains a significant challenge, with most scores below 3 on a 5-point scale. This is inline with  observations reported in~\citep{Bernard:2025:SIGIR-AP,Bernard:2025:WSDM}.
However, a closer look at the table reveals some interesting findings. 

First, the simulators demonstrate significant disagreement on system ranking. For instance, on the MovieBot dataset, ABUS ranks IAI MovieBot last for user satisfaction (1.5), while LLM-DP ranks it first (2.5). Furthermore, even when simulators agree on the top-performing system (e.g., IAI MovieBot for Recommendation Relevance on ReDial), they often diverge on the magnitude of its performance (a 4.7 from ABUS vs. 2.4/2.8 from the LLM simulators).

The simulators also show distinct characteristics. ABUS appears to be the ``most opinionated,'' assigning both the highest score in the table (4.7) and some of the lowest (1.0). The LLM simulators, by contrast, operate within a more compressed range, rarely awarding scores above 3.3. Moreover, the LLM simulators are not interchangeable. Despite producing similar results at times, LLM-DP consistently yields equal or higher average scores than LLM-SP on the INSPIRED dataset across all three metrics.

Finally, CRS performance itself is highly dataset-dependent. Some systems show extreme variability; for example, IAI MovieBot's recommendation relevance (via ABUS) peaks at 4.7 on ReDial---the highest score in the table---but drops to a mediocre 2.1 on the other datasets. In contrast, other systems demonstrate high consistency. BARCOR\_ReDial's fluency score from ABUS, for instance, remains stable and high across all three datasets (3.3, 3.5, 3.4).

%However, the results indicate that conversations between ABUS and IAI MovieBot tend to be less satisfying than with other CRSs.
% This suggests that the user simulators do not identify strong differences between the CRSs.

\paragraph{Enabled Research Directions.}
Considering the results and the synthetic dialogues generated, we can envision different research questions that UserSimCRS v2 can help investigate. For example, these include studying the influence of different user simulator types (e.g., agenda-based vs. LLM-based) and configurations (e.g., different LLMs, different prompts, datasets, and interaction models) on the evaluation outcomes, comparing the performance of different CRSs, and analyzing the synthetic dialogues with regard to their discourse structure, characteristics, or breakdowns. These are merely illustrative of the much broader scope of research that UserSimCRS v2 enables; however, the focus of this paper remains on the presentation of the toolkit itself.

\section{Conclusion}
\label{sec:concl}

In this work, we presented an upgraded version of UserSimCRS, designed to facilitate a more comprehensive and flexible comparison of user simulators and conversational recommender systems through simulation-based evaluation. 
This new version aligns the toolkit with current research by enhancing the classic agenda-based user simulator with LLM-powered components, introducing LLM-based simulators, adding new automatic evaluation utilities, and providing support for a wider range of CRSs and datasets.

Future work will focus on enhancing the integration of user modeling components to better support the simulation of diverse user populations and exploring novel evaluation metrics. Furthermore, we aim to continue reducing the technical entry barrier to encourage widespread adoption of the toolkit by the community, thereby fostering research on user simulation.

\small{\subsubsection{\ackname} This research was partially supported by the Norwegian Research Center for AI Innovation, NorwAI (Research Council of Norway, project number 309834), and by the German Federal Ministry of Education and Research and Joint Science Conference under the PLan\_CV project (reference number 03FHP109).}

\small{\subsubsection{Disclosure of Interests.}
The authors have no competing interests to declare that are relevant to the content of this article.}

\bibliographystyle{splncs04nat}
\bibliography{ecir2026-usersimcrs.bib}

\end{document}